\documentclass[showpacs,aps,prb,nofootinbib,nobibnotes,superscriptaddress,floatfix,preprint]{revtex4-1}

\usepackage{amssymb}
\usepackage{graphicx,amsmath}
\usepackage{bm}
\usepackage{times}
\usepackage{epstopdf}
\usepackage{xcolor}
\def\expect#1{\left\langle #1 \right\rangle}
\def\hc{\text{h.c.}}
\def\OmD{\bar{\Omega}}
\def\Ome{\check{\Omega}}
\def\fD{\bar{f}}
\def\fe{\check{f}}

\begin{document}

\title{Controlling the energy gap of a tunable two-level system by ac drive}
\date{\today }
\author{G. Oelsner}
\email{gregor.oelsner@leibniz-ipht.de}
\affiliation{Leibniz Institute of Photonic Technology, P.O. Box 100239, D-07702 Jena, Germany}

\author{U. H\"ubner}
\affiliation{Leibniz Institute of Photonic Technology, P.O. Box 100239, D-07702 Jena, Germany}
\author{E.~Il'ichev}
\affiliation{Leibniz Institute of Photonic Technology, P.O. Box 100239, D-07702 Jena, Germany}
\affiliation{Novosibirsk State Technical University, 20 Karl Marx Avenue, 630092 Novosibirsk, Russia}

\begin{abstract}
We study the influence of a strong off-resonant driving signal to the energy levels of a superconducting flux qubit both experimentally and theoretically. In the experiment, we carry out a three-tone spectroscopy. This allows us to directly observe the modification of the qubit's energy levels by the dynamical Stark shift caused by the driving signal. A theoretical treatment including corrections from both, rotating and counter-rotating frame, allowed us to completely explain the observed experimental results and to reconstruct the influence of the strong driving to the dissipative dynamics as well as to the coupling constants of the qubit. As one potential application, the tunability of the minimal energy-level splitting of a superconducting qubit by a microwave induced dynamical Stark shift can help to overcome the parameter spread induced by the micro fabrication of superconducting artificial quantum circuits.
\end{abstract}

\maketitle

\section*{Introduction}
Several unique properties of superconducting devices make them promising platforms for modern quantum technologies. Due to their macroscopic size, superconducting quantum circuits exploit large coupling constants and deliver a good scalability as well as controllability.  While first experiments \cite{Martinis1987, Nakamura1999, Born2004, Ilichev2004, Wallraff2004, Astafiev2010} were mainly focused on the demonstration of quantum effects, they were later used for the implementation of quantum gates \cite{Plantenberg2007, Fedorov2011}. Meanwhile complex circuits are designed and operated in context of quantum computation \cite{Devoret2013, OMalley2016, Wendin2017, Arute2019} and, for example, in such promising directions as quantum metamaterials \cite{Macha2014, Shulga2018}.

Besides decoherence \cite{Martinis2005, Ithier2005}, parameter spread and a certain lack of reproducibility of the superconducting quantum circuits due to micro-fabrication tolerances remain severe challenges \cite{Oelsner2017b}. In this context, additional tunabilities were discussed and realized. For instance, a so-called $\alpha$-loop \cite{Mooij1999, Fedorov2011a} has been introduced to flux qubits. Although this allows using the flux qubit at its degeneracy point with an energy splitting as desired, any additional circuit is a potential source of low-frequency noise and thus introduces more decoherence \cite{Esteve1986,Ithier2005,Yoshihara2006,Neeley2008}. In order to avoid such unwanted effects, we propose to use a natural tunability inherent in any addressable quantum system, namely the radiative shift of energy levels. By making use of a flux qubit as one example of solid-state quantum systems, we demonstrate the shift of its energy levels by off-resonant driving. This effect can be interpreted as ac Zeeman shift that is the magnetic equivalent to the ac Stark shift, sometimes also called dynamic Stark shift. The latter has been already observed in early circuit-QED experiments \cite{Schuster2005}. While the energy level shift induced by a single frequency driving tone can be accurately described in frame of Floquet theory \cite{Shirley1965,Son2009,Tuorila2010,Koski2018}, its influence onto coupling constants with resonators or further driving tones and to dissipative dynamics needs to be as well considered.

In this paper, we analyze in detail the usability of such an ac field-induced shift to tune the energy-level splitting of a two-level system. We perform spectroscopic measurements on a flux qubit to determine the shifted transition frequency. Additionally, we develop a theoretical model for a driven tunable quantum two-level system that explains the observed measurement results. Especially, our investigations enable to calculate the influence of the off-resonant drive to the dissipative dynamics of the manipulated qubit and its coupling to resonators and further microwave fields. Additionally, our method is important for other applications of dynamical level shifts, for example, to improve the accuracy of a suggested microwave power calibration technique \cite{Schneider2018}.

\section{Experimental characterization}
\begin{figure}[htb]
\includegraphics[]{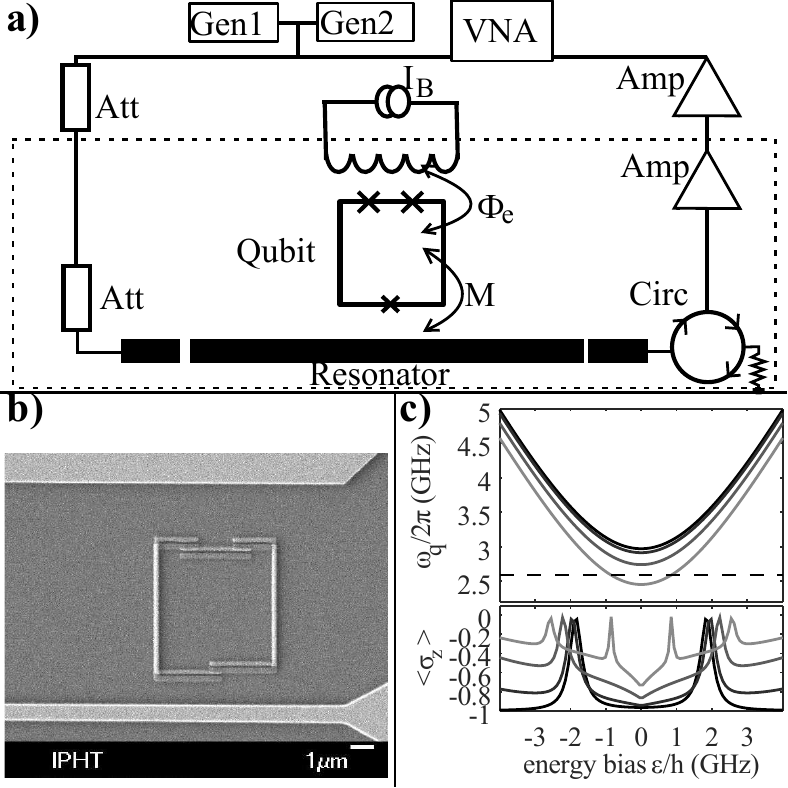}
\caption{a) Schematic of the experimental setup. b) SEM image of the resonator's central part with the coupled flux qubit c) Calculated qubit energies and population as a function of energy bias. From dark to light color the driving amplitude $\Omega_d/2\pi$ is increased from 0 to 3~GHz in steps of 1~GHz while its frequency $\omega_d$ is chosen at the resonators third harmonic. The frequency of the spectroscopic signal $\omega_s/2\pi$ is fixed at 3.5~GHz with an amplitude $\Omega_s/2\pi$ of 10~MHz. The dashed line indicates the resonator's fundamental mode frequency. } \label{Fig:sample}
\end{figure}

We used an experimental setup as sketched in Fig.~\ref{Fig:sample}. The sample is mounted to the milliKelvin-stage of a dilution refrigerator (indicated by the dashed line). The probing tone of a vector network analyzer (VNA) is combined with two additional signals from RF-Generators (Gen). The resulting signal is transferred into the cryostat. Attenuators (Att) at room temperature, 1.5~K, and 20~mK with 20 dB attenuation each are used to thermalize the signal before it drives a coplanar waveguide resonator and thereby also the qubit. The transmitted signal is amplified by a cryogenic amplifier located at the 4 Kelvin stage with a noise temperature of about 7~K together with several room temperature amplifiers before eventually measured by the VNA. An isolator (Circ), located at the refrigerator base, shields the noise of the cold amplifier from the sample. A small Helmholtz coil system is supplied with a dc-bias current ($I_B$) to apply a flux bias to the qubit. The used twisted pair lines are filtered by RC and copper-powder filters at 1.5~K and 20~mK, respectively, together with feedthrough filtering at room temperature. To reduce the influence of magnetic field noise, two mu-metal and one superconducting lead shields enclose the sample.

The central element of the experiment is a flux qubit that consists of a $5\times5$~$\mu$m$^2$ superconducting loop interrupted by three Josephson junctions. It is fabricated by standard two-angle shadow evaporation of aluminum with an intermediate oxidation step \cite{Oelsner2017b}. The qubit is placed in the gap between the central conducting line and the ground plane of a coplanar resonator as shown in the SEM image of Fig.~\ref{Fig:sample}~b). The resonator structure is formed by reactive ion etching of a thin niobium film sputtered on a high-ohmic silicon substrate. Its length is chosen for a fundamental mode frequency of $\omega_r/2\pi = 2.59$~GHz, and a quality factor of $12\times 10^4$ was experimentally determined. The chosen geometry results in a mutual inductance between qubit and resonator of about 1.2~pH.

In the experiment, we in general apply three signals to the resonator. One strong driving signal at the resonator's third harmonic is used to dress the qubit states. We have chosen the third harmonic because its coupling to the qubit is well defined allowing to relate the input power in dBm to the driving amplitude of the qubit (see below). By dressing the qubit, its effective level splitting and dissipative rates are modified\cite{Shevchenko2014}, and amplification and lasing could in principle be achieved depending on experimental control parameters \cite{Oelsner2013, Neilinger2015, Neilinger2016}. The corresponding level shift is exemplarily shown in Fig.~\ref{Fig:sample}~c). A second tone with variable frequency is used for spectroscopy of the qubit's shifted energy level difference. When the resonance condition is met, the qubit population is modified as shown in the lower plot of Fig.~\ref{Fig:sample}~c). This leads to a partly reduction of the dispersive shift \cite{Greenberg2002,Oelsner2010} induced to the resonator by the qubit. A third weak probe beam is applied close to the resonators fundamental mode for probing this frequency shift in a transmission-type measurement. An analysis of its variation as a function of the applied driving and spectroscopic signals' amplitudes and frequencies gives information of the dressed qubit's properties. Corresponding measurement results are shown in Fig.~\ref{Fig:spec}.

\begin{widetext}
\begin{center}
\begin{figure}[bht]
\includegraphics[]{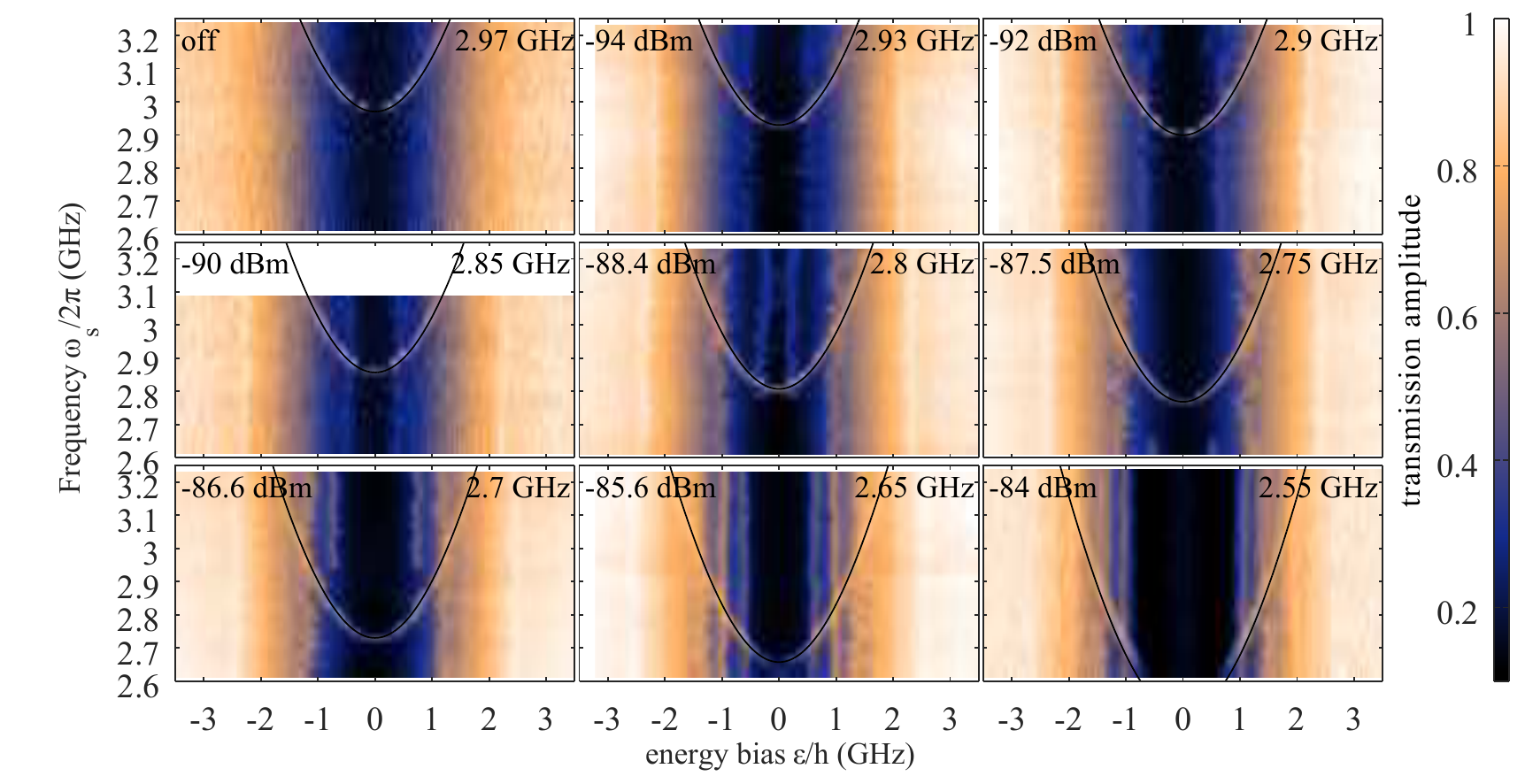}
\caption{(Color online) Spectroscopy of the dressed qubit levels. The normalized transmission amplitude at the resonators fundamental mode frequency $\omega_r$ is plotted as a function of the energy bias of the qubit $\varepsilon$ and the frequency of the spectroscopy signal $\omega_s$. For the different sub-figures the amplitude of the dressing signal at the third harmonic of the resonator is increased, while for the upper left plot no additional driving is applied. A horizontal cut through any of the figures reveals the dispersive shift of the resonator frequency by the curvature of the qubit's ground state\cite{Greenberg2002}. This shift is canceled if the spectroscopic frequency $\omega_s$ is resonant to the qubit. Accordingly, its level splitting as a function of the energy bias is observed. With increasing dressing amplitude $\Omega_d$ the qubit gap is effectively reduced. The extracted value of this minimal level splitting in each respective plot is given in units of GHz together with the input driving power in dBm applied to the sample by the black text. The black lines correspond to the calculated qubit level splitting as found from Eq.~\eqref{Eq:ac-shift} with the same numbers as used for Fig.~\ref{Fig:Calc}.} \label{Fig:spec}
\end{figure}
\end{center}
\end{widetext}

Therein, each subplot represents a spectroscopy of the qubit's dressed energy levels for a certain amplitude of the driving signal. From the spectroscopic curve without the strong dressing signal (upper left plot in Fig.~\ref{Fig:spec}) the parameters of the qubit can be extracted. The minimal level splitting (energy gap) and persistent current are found as $\Delta/h = 2.97$~GHz and $I_p = 160$~nA, respectively. Here, $h$ denotes the Planck constant. Due to parameter spread in the fabrication, the experimental determined qubit parameters are each about 15~\% off the aimed values. The persistent current connects the energy bias $\varepsilon = 2 I_p \left(\Phi_0/2-\Phi_e\right) $ to the external bias flux $\Phi_e$. The mutual inductance and the currents in resonator $I_r$ and qubit $I_p$ lead to an interaction energy $\hbar g =  M I_p I_r $. It can be expressed by the coupling constant of qubit and resonator's fundamental mode $g/2\pi$ that takes a value of about 3~MHz.

While the experimental measured spectroscopic line stays qualitatively the same, with increasing driving amplitude the effective gap frequency of the qubit is reduced. The extracted gap values are given in Fig.~\ref{Fig:spec} as black text in GHz. As it will be shown below, the observed reduction can be explained by a strong dynamical Zeeman shift of the qubit's energy levels. We note that this is true only away from the resonance point. When considering small qubit-drive detunings (compared to the driving induced splitting) the description of the system should be carried out in the dressed-state basis \cite{Oelsner2013}. The latter allows for example the discussion of amplification and damping of the resonator transmission by the interaction of the dressed-states with the fundamental mode.

\section{Theoretical description}

In the experimental situation presented above, three signals with different amplitudes and frequencies are applied to a qubit-resonator system. A corresponding Hamiltonian in the eigenbasis of the qubit reads \cite{Hauss2008a,Oelsner2010,Shevchenko2014}
\begin{equation}\label{Eq:Ham_full}
\begin{split}
H &= \hbar \omega_r a^\dagger a+ \frac{\hbar \omega_q}{2} \sigma_z+ \hbar \Omega_p \left[ e^{-i\omega_pt}a^\dagger + e^{i\omega_pt} a\right] \\
& \!+ \!\left( \!\frac{\varepsilon}{\omega_q} \sigma_z \! + \! \frac{\Delta}{\omega_q} \sigma_x \right)\left[\!\sum_{i = d,s} \! \Omega_i \cos \omega_i t \! + \! g \left(a+a^\dagger\right) \! \right].
\end{split}
\end{equation}
It includes the quantized fundamental mode with frequency $\omega_r$ represented by the creation and annihilation operators $a^\dagger$ and $a$. The qubit is discussed in a two-level approximation. Therefore, its dynamics can be described with the Pauli operators $\sigma_i$. The energy splitting between the qubit eigenstates is given by $\hbar \omega_q = \sqrt{\Delta^2+\varepsilon^2}$. The probing, driving, and spectroscopic signals with amplitudes $\Omega_i$ and frequencies $\omega_i$ for the respective indices $p$, $d$ and $s$ are included as quantum (probing) and classical (driving and spectroscopic) signals, respectively. Thus the coherent coupling between the fundamental mode and the qubit with coupling energy $\hbar g$ is explicitly denoted. Because we are interested in calculating its transmission, the probe tone is applied to the resonator while the other two are considered to drive the qubit. To shorten the equations, we indicate in the following the respective multiplication with $\Delta/\hbar\omega_q$ and $\varepsilon/\hbar\omega_q$ by a bar and a hat down.

%Note, by neglecting the externally applied signals $\Omega_i = 0$ and for large detuning, $\delta_{qr} = \omega_q - \omega_r \gg g$, between resonator and qubit frequency the system Hamiltonian can be brought into the illustrative form\cite{Blais2004,Oelsner2010}
%\begin{equation}\label{Eg:disp-regime}
%H \approx \hbar\left( \omega_r + \frac{{g^\prime}^2}{\delta_{qr}} \sigma_z \right) a^\dagger a + \hbar \left( \omega_q + \frac{{g^\prime}^2}{\delta_{qr}} \right) \sigma_z .
%\end{equation}
%Different interpretations of this equation for the dispersive regime are possible. For instance, the term $g^2_\Delta/\delta_{qr}a^\dagger a \sigma_z$ can be interpreted either as the ac-Zeeman shift of the qubit's eigenfrequency or as the dispersive shift of the oscillator frequency as a function of the qubit state. The latter interpretation allows an estimation of the transmitted probing signal, simply by analyzing the influence of the strong signals in \eqref{Eq:Ham_full} to the mean qubit's population $\expect{\sigma_z}$.

A complete theoretical description of the dynamics has to include the dissipation. We use the Lindblad operators $L_q(\rho)$ including qubit relaxation and pure dephasing with the respective rates $\Gamma_r$ and $\gamma_\varphi$ as well as $L_r(\rho)$ for the resonators photon decay with rate $\kappa$. This assumes that the noise causing dissipation can be considered as Markovian and that the qubit's dynamics can be described by relaxation time $T_1 = 2\pi /\Gamma_r$ and coherence time $T_2 = 4\pi / (2\gamma_\varphi+ \Gamma_r)$. Besides being a good model for atomic systems \cite{Oelsner2019a}, this is also a good approximation for a flux qubit close to the degeneracy point \cite{Yoshihara2006}. Thermal excitations of the qubit and the resonator can be neglected compared to the relaxation because of the low sample temperature at about $20$~mK. With denoting the systems density matrix by $\rho$, they read in the qubit eigenbasis
\begin{equation} \label{Eq:Linblad}\begin{split}
L_q(\rho) &= \frac{\Gamma_r}{2} \left( 2\sigma_- \rho \sigma_+ - \left\{ \sigma_+ \sigma_- , \rho \right\} \right) +\frac{\gamma_\varphi}{2} \left( \sigma_z \rho \sigma_z - \rho \right) \\
L_r(\rho) &= \frac{\kappa}{2} \left( 2a\rho a^\dagger - \left\{ a^\dagger a , \rho \right\} \rho \right).
\end{split}\end{equation}
Here the anti-commutator is written as $\left\{ A , B \right\} = AB + BA$, and $\sigma_\pm = 1/2 (\sigma_x \pm i\sigma_y)$ denote the qubit raising and lowering operators.

\subsection*{Qubit level shift by the driving}
\label{Theo1}
In a first step, we analyze the level shift introduced by the strong driving signal to the qubit. It is given by the time average of the interaction energy expressed as the product of the induced dipole moment
\begin{equation}
p(t) = \expect{\sigma_+} +\expect{\sigma_-}
\end{equation}
with the applied field \cite{LeKien2013} resulting in an energetic shift of
\begin{equation}\label{Eq:energyshift}
\partial E = \hbar \OmD_d \overline{\cos\omega_d t \cdot p(t)}.
\end{equation}
Neglecting the probing and spectroscopic signal as well as the diagonal driving $\propto \sigma_z$ in \eqref{Eq:Ham_full} results in the Hamiltonian of a driven quantum two-level system
\begin{equation}
H_d = \frac{\hbar \omega_q}{2} \sigma_z + \hbar\OmD_d \cos \omega_d t \sigma_x.
\end{equation}
Its time evolution is given by Bloch equations
\begin{equation}\label{Eq:Bloch1}\begin{split}
\expect{\dot{\sigma}_z} =& - \Gamma_r\left( \expect{\sigma_z}+1\right) - i \OmD_d  \cos\omega_dt\left[ \expect{\sigma_+} - \expect{\sigma_-} \right],\\
\expect{\dot{\sigma}_+} =& i \left(\omega_q  + i \Gamma_\varphi \right) \! \expect{\sigma_+} - i \frac{\OmD_d }{2}\! \left(  e^{-i \omega_dt}\! + \! e^{i \omega_d t} \right) \! \expect{\sigma_z}, \\
\expect{\dot{\sigma}_-} =&  \expect{\dot{\sigma}_+}^*.
\end{split}\end{equation}
Above, $\Gamma_\varphi = \Gamma_r/2 + \gamma_\varphi$ is the qubit's decoherence rate and $^*$ stands for complex conjugate. To solve these equations, we replace
\begin{equation}\label{Eq:Ansatz}
\expect{\sigma_+} = s_+ e^{-i\omega_d t} +s_- e^{i\omega_d t}
\end{equation}
in the second equation of \eqref{Eq:Bloch1} and find
\begin{equation}\label{Eq:start}\begin{split}
s_+ &= \frac{\OmD_d }{2 \left[\omega_q+\omega_d - i\Gamma_\varphi \right]} \expect{\sigma_z}, \\
s_- &= \frac{\OmD_d }{2 \left[\omega_q-\omega_d - i\Gamma_\varphi \right]} \expect{\sigma_z}.
\end{split}\end{equation}

These solutions enable calculating the mean population of the qubit from the steady state ($\expect{\dot{\sigma_z}}=0$) of the first line of \eqref{Eq:Bloch1}. Neglecting oscillating terms it is
\begin{equation}\label{Eq:expectsigz1}
\expect{\sigma_z} = \frac{-\Gamma_r}{\Gamma_r + \frac{\OmD_d^2}{2} \Im\left(\frac{1}{\omega_q-\omega_d-i\Gamma_\varphi} +\frac{1}{\omega_q+\omega_d-i\Gamma_\varphi}\right)}.
\end{equation}
From above result, we conclude that the qubit stays basically in the ground state if the driving signal is far detuned from resonance so that $\delta_\pm = \omega_q \pm \omega_d$ is much larger than the driving amplitude $\Omega_d$. The $\pm$ sign accounts for the respective shifts obtained in the rotating and counter-rotating frame, as we demonstrate below. The off-resonant driving limit allows to set $\expect{\sigma_z} = -1$ when calculating the induced dipole moment, that we rewrite to
\begin{equation}
p = \left( \tilde{p} e^{-i\omega_d t}+\tilde{p}^* e^{i\omega_d t} \right)
\end{equation}
with $\tilde{p} = s_+ + s_+^*$. Finally the energy shift is found to be
\begin{equation}\label{Eq:ac-shift}\begin{split}
\partial E =& \frac{\hbar\OmD_d^2}{2} \Re \left(\frac{1}{\omega_q-\omega_d+i\Gamma_\varphi} +\frac{1}{\omega_q+\omega_d-i\Gamma_\varphi}\right)\\
\approx & \hbar\OmD_d^2 \frac{\omega_q}{\omega_q^2-\omega_d^2} = \hbar \omega_{ac}.
\end{split}\end{equation}
Above equation describes the shift of the energy levels with contributions of the total dynamical Zeeman shift and the Bloch-Siegert shift. For the approximation in the lower line, we considered large the detunings $\delta_\pm$ compared to the decoherence rate $\Gamma_\varphi$. The equation is consistent with the second order shift found by Floquet theory \cite{Rosenbusch2009}. It also very well describes the observed experimental shift of the qubit levels by the off-resonant driving, as demonstrated by the black lines in Fig.~\ref{Fig:spec}. We note that both terms in \eqref{Eq:ac-shift} contribute in our experiment at about same order of magnitude.

\subsection*{Rescaling of coupling and dissipation}
Above we followed an approach often applied to natural atomic systems to accurately describe their level shift induced by off-resonant driving fields. It also includes the effect from the linewidth broadening due to decoherence. In contrast to the tunable quantum system that represents a flux qubit, there exist no diagonal ($\sigma_z$) coupling between atoms and applied signals in first order. Moreover, it is not clear what influence the interaction responsible for shifting the energy levels has on the coupling coefficients to both, fundamental mode and spectroscopy signal, since a re-normalization of the coupling coefficients can be expected.

In order to clarify these issues, we performed calculations to map back the energy shift found in the rotating frames with $\pm \omega_d$ to the laboratory frame, see appendix \ref{App}. As shown there, we eliminated the off-diagonal driving with $\OmD_d$ by a transformation $U = U_t U_+ U_t^\dagger U_t^\dagger U_- U_t$, with $U_t = e^{i\omega_d t \sigma_z /2}$ and $U_\pm = e^{i\sigma_y \arctan\left(\OmD_d/\delta_\pm\right)/2}$. Because the rotations around the y-axis are assumed small for $U_\pm$, these transformations are expanded to second order in $\OmD_d^\prime / \delta_\pm$. An analog approach has been used to calculate the Bloch-Siegert shift in the dressed-state basis for a quantized radiation field coupled to a two-level system \cite{FornDiaz2010,Naderi2011}.

Additionally, we shift the time dependence introduced by the probing signal to the coupling terms with the unitary transformation $U_p = e^{i\omega_p a^\dagger a}$. Note that this basis change only effects terms containing fundamental mode operators. Transformed in this way and in rotating wave approximation Hamiltonian \eqref{Eq:finalHam2} reads
\begin{equation} \label{Eq:Ham_firststep}
\begin{split}
\widehat{H} =& U HU^\dagger \\
&\hbar \frac{\Omega_R}{2} \sigma_z + \hbar \delta_{rp} a^\dagger a + \hbar \Omega_p \left[ a^\dagger+ a\right]\\
&\! \hbar\left[ \Omega_s \cos\omega_st +g \left[e^{i\omega_p t} a^\dagger + \hc \right] \right) \left( \alpha \sigma_z - \beta \sigma_x \right)
\end{split}
\end{equation}
where the detuning $\delta_{rp}$ between resonator and qubit was introduced and $\hc$ stands for complex conjugate. The qubit frequency splitting
\begin{equation}\label{Rabi_freq}
\Omega_R = \sqrt{\left( \omega_q + \omega_{ac} \right)^2 + \left(\frac{2\varepsilon}{\Delta} \omega_{ac}\right)^2 }
\end{equation}
includes the complete dynamical Zeeman shift for large detuning (see above). For small detunings $\delta_-$ the system is more conveniently described in the dressed state basis \cite{Shevchenko2014,Oelsner2013}. The last term in \eqref{Rabi_freq} $\varepsilon\omega_{ac}/\Delta$, resulting from the diagonal coupling, is a direct consequence of the qubit tunability. Because this small term vanishes for natural (un-tunable) quantum systems, its influence in a driven system's dynamics has not been discussed in literature.

The resulting re-normalization of the coupling in \eqref{Eq:Ham_firststep} is represented by the dimensionless factors $\alpha = \frac{\varepsilon}{\hbar\omega_q\Omega_R}\left( A \left[ \omega_q + \omega_{ac} \right] + 2B \omega_{ac} \right)$ and $\beta = \frac{\Delta}{\hbar\omega_q\Omega_R} \left( B \left[ \omega_q + \omega_{ac} \right]+ \frac{2\varepsilon^2}{\Delta^2} A \omega_{ac}\right)$. These factors depend on the variables $A$ and $B$ which are of second order in $\Omega_d/\delta_\pm$, see appendix~\ref{App}. We conclude that for small driving amplitudes compared to the detuning, the modification in coupling constants is small compared to the first-order level shift. Since this change in coupling results from the modification of the qubit's dipole moment, a similar effect can be expected for its interaction with the environment, and only second order corrections are expected in the dissipative rates.

In appendix~\ref{App2} we present calculations to determine the modifications of the Lindblad operator of the flux qubit in presence of a strong off- resonant drive. We find
\begin{equation}\begin{split}\label{Eq:Linbladmod}
\widehat{L}(\widehat{\rho}) &= \frac{\widehat{\Gamma}_r}{2} \left( 2\sigma_- \widehat{\rho} \sigma_+ - \left\{ \sigma_+ \sigma_- , \widehat{\rho} \right\} \right) \\
&+ \frac{\widehat{\Gamma}_e}{2} \left( 2\sigma_+ \widehat{\rho} \sigma_- - \left\{ \sigma_- \sigma_+ , \widehat{\rho} \right\} \right) +\frac{\widehat{\gamma}_\varphi}{2} \left( \sigma_z \widehat{\rho} \sigma_z - \widehat{\rho} \right)
\end{split}\end{equation}
with the respective relaxation, excitation, and dephasing rates
\begin{equation}\begin{split}\label{Eq:modrates}
\hat{\Gamma}_r &=   \Gamma_r - \frac{C}{2}\left( \Gamma_r - \gamma_\varphi \right) ,  \\
\hat{\Gamma}_e &= \frac{C}{2} \gamma_\varphi ,\\
\hat{\gamma}_\varphi &= \gamma_\varphi + \frac{C}{2}\left( \Gamma_r - 2\gamma_\varphi \right).
\end{split}\end{equation}
Above we shortened $C = \OmD_d^2\left[ 1/\delta_-^2+ 1/\delta_+^2 \right]$. In interpretation, the mixing of the states induced by the strong driving signal leads to a modification of the dissipative dynamics. This is required to account for the slight change in population of the ground and excited states. Note that compared to the rates in the dressed state basis \cite{Hauss2008a,Wilson2010}, here, the lower populated state is always the higher energetic state. Therefore, effects connected to Rabi-resonance, such as amplification or lasing in strongly driven systems \cite{Grajcar2008, Oelsner2013, Neilinger2015} cannot be explained in this corrected laboratory frame. Additionally, the corrections are of second order and thus can be neglected for small driving amplitudes compared to the detunings $\delta_\pm$. Interestingly, the rate of pure dephasing is reduced for strong driving signals which indicates effects used in dynamical decoupling schemes, when the decoherence is dominated by pure dephasing.

\subsection*{Spectroscopy of the shifted qubit levels}
Hamiltonian \eqref{Eq:Ham_firststep} describes a driven two-level system coupled additionally to the mode of a resonator and tunable by both the dc-bias field and the power of an additional driving signal. We are interested in resonances between the shifted qubit levels and the spectroscopic signal that we probe by the resonator.

For the system given by Hamiltonian~\eqref{Eq:Ham_firststep} together with qubit dissipation \eqref{Eq:Linbladmod} and resonator photon decay \eqref{Eq:Linblad} we can write Maxwell-Bloch equations for the expectation values of qubit and resonator operators. When neglecting the diagonal coupling terms and keeping only resonant terms ($a^\dagger \sigma_-$ and $a \sigma_+$) they read
\begin{align}
\dot{\expect{a}} =& -i \left( \delta_{rp} - i\frac{\kappa}{2} \right) \expect{a} -i\Omega_p - i g \beta \expect{\sigma_-}e^{i\omega_p t}, \label{Eq:MaxBloch1}\\
\dot{\expect{\sigma_-}} =& -i \left(\Omega_R - i \hat{\Gamma}_\varphi \right)\expect{\sigma_-} + i \beta g\expect{a \sigma_z}e^{-i\omega_pt} \nonumber\\
& + i \beta \Omega_s \cos\omega_s t \expect{\sigma_z} ,\label{Eq:MaxBloch2}\\
\dot{\expect{\sigma_z}} =& -4 \beta \Im\left( g \expect{a^\dagger \sigma_-}e^{i\omega_p t} + \Omega_s \cos\omega_s t  \expect{\sigma_-} \right) \nonumber \\
&- \hat{\Gamma}_r \left( 1+\expect{\sigma_z} \right)+ \hat{\Gamma}_e \left( 1-\expect{\sigma_z} \right) .\label{Eq:MaxBloch3}
\end{align}
In the following, we assume that qubit and resonator expectation values factorize $\expect{a\sigma_i} = \expect{a}\expect{\sigma_i}$ meaning that qubit and resonator are not entangled. We use this standard practice of quantum optics because we consider coherent driving and continues measurements allowing the use of the semi-classical approach \cite{Shevchenko2014}. Furthermore, as shown below, the analytical solution found under this assumption agrees well to our experimental results. Motivated by section \ref{Theo1} we use the ansatz
\begin{equation}
\expect{\sigma_-} = s_p e^{-i\omega_p t} + s_s e^{-i\omega_s t},
\end{equation}
that in Eq.~\eqref{Eq:MaxBloch2} results in
\begin{equation} \begin{split}
s_p &= \frac{\beta g}{\Omega_R - \omega_p -i \hat{\Gamma}_\varphi} \expect{\sigma_z} \expect{a},\\
s_s &= \frac{\beta \Omega_s}{2\left(\Omega_R - \omega_s -i \hat{\Gamma}_\varphi\right)} \expect{\sigma_z}.
\end{split}\end{equation}
Here and in the following we only keep stationary terms. With these expressions we calculate the stationary solution of \eqref{Eq:MaxBloch1} as well as \eqref{Eq:MaxBloch3} and find
\begin{equation}\label{Eq:expecta}
\expect{a} = \frac{-\Omega_p}{\delta_{rp}- i \frac{\kappa}{2} + \frac{g^2\beta^2}{\Omega_R-\omega_p - i \hat{\Gamma}_\varphi}\expect{\sigma_z}},
\end{equation}
\begin{equation}\label{Eq:pop}
\expect{\sigma_z} = \frac{\hat{\Gamma}_e-\hat{\Gamma}_r}{\hat{\Gamma}_e+\hat{\Gamma}_r - \beta^2\hat{\Gamma}_\varphi \! \left(\frac{4g^2 \expect{a}\expect{a^\dagger}}{\left[ \Omega_R - \omega_p \right]^2 +\hat{\Gamma}_\varphi^2} \! + \! \frac{\Omega_s^2}{\left[ \Omega_R - \omega_s \right]^2 +\hat{\Gamma}_\varphi^2} \right) }.
\end{equation}
Therewith, the transmission through the resonator is given by $\cite{Oelsner2010}$
\begin{equation}
\label{Eq:res-freq}
t = \frac{i}{2} \frac{\kappa}{\delta_{rp}- i \frac{\kappa}{2} + \frac{g^2\beta^2}{\Omega_R-\omega_p  - i \hat{\Gamma}_\varphi} \expect{\sigma_z}}.
\end{equation}

The transmission coefficient above describes a resonance curve around the cavities fundamental-mode frequency. Still, by the coupling to the qubit, this resonant frequency becomes a function of the qubit population $\expect{\sigma_z}$ as well as the detuning of resonator and effective qubit frequency $\Omega_R-\omega_p$. The latter is reduced by stronger off-resonant driving amplitudes $\Omega_d$, consequently increasing the qubit's pull of the cavity resonance. On the other hand, the qubit population, as given in the lower line of Eq.~\ref{Eq:pop}, can be influenced by the spectroscopic signal as well as the signal in the cavity. The strength of a qubit excitation by these signals depends on their respective detuning from the qubit's effective level splitting allowing for spectroscopic measurements. Namely, this relation enables to determine the shifted qubit levels as well as the observation of cavity-qubit resonances.

By the strong off-resonant driving in our experiment, we reduce the detuning between resonator's fundamental mode and effective qubit significantly. Therefore, the corresponding term in Eq.~\ref{Eq:pop} cannot be omitted. However, the effect of the resonator's fundamental-mode signal to the qubit population can be assumed small for any practical detuning since the system is operated in the intermediate coupling regime where $g < \Gamma_\varphi$ \cite{Oelsner2010} and with small probing powers. Therefore away from resonance, the actual mean photon number $N$ is not of critical importance, as long as it is small enough not to produce a significant change of the qubit population. On the other hand in resonance $\Omega_R = \omega_p$, a change in photon number strongly influences to the qubit population and vice versa. For the theoretical simulations, we thus assume a constant photon number $N$ in the fundamental mode that would correspond to the value in resonance and is smaller than the rather insignificant off-resonant one. It enters to the calculations by the product of expectation values in Eq.~\eqref{Eq:pop} $N = \expect{a}\expect{a^\dagger}$ considering that the driving produces a coherent state in the resonator.

\begin{widetext}
\begin{center}
\begin{figure}[htb]
\includegraphics[width = 16 cm]{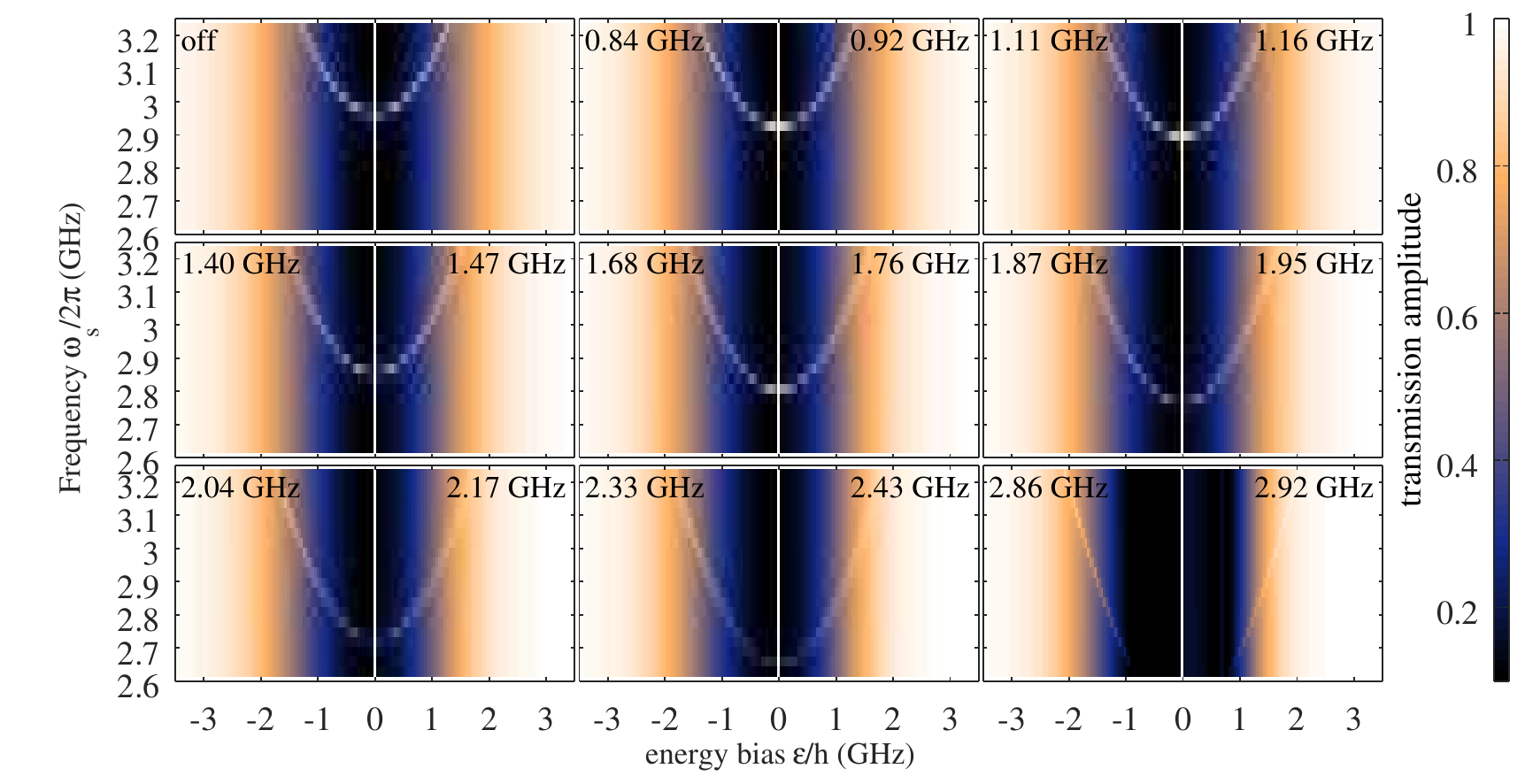}
\caption{(Color online) Calculated transmission amplitude as a function of the energy bias $\varepsilon$ and the frequency of the spectroscopic signal $\omega_s/2\pi$. From up left to low right the amplitude of the dressing signal $\Omega_d$ is increased as in Fig.~\ref{Fig:spec}. The inserted numbers on the left and right of each subplot correspond to the driving amplitude $\Omega_d$ that is used for the calculation and to the one reconstructed from the applied power following Eq.~\ref{Eq:powertodrive}, respectively. The first is defined by the reconstructed effective, gaps see Fig.~\ref{Fig:spec}. Additionally, we used only the first order corrections to the dynamics for simulation of the negative bias region, while we account for the modification of dissipation and coupling to second order on the positive bias data points. This separation is indicated by the inserted white line at zero bias.} \label{Fig:Calc}
\end{figure}
\end{center}
\end{widetext}

With the use of Eq.\eqref{Eq:res-freq} we simulate our experimental findings as demonstrated in Fig.~\ref{Fig:Calc}. As parameters for the simulation we used an amplitude of the spectroscopic signal $\Omega_s = 1$~MHz, a photon number in the fundamental mode of $N = 5$ photons, as well as a respective relaxation and pure dephasing rate $\Gamma_r = 10$~MHz and $\gamma_\varphi = 20$~MHz. The values for the driving amplitude are given as black text in each of the subplots of Fig.~\ref{Fig:Calc}. For comparison, always the second of the two numbers is reconstructed from the input driving power as
\begin{equation}\label{Eq:powertodrive}
\Omega_d = 4 g_d \sqrt{N_d} = 4g_d\frac{1}{\kappa_d}\frac{C_c}{2} \sqrt{\frac{10^{(P_{\text{in}}-30)/10}}{Z}} \sqrt{\frac{h \omega_d}{C_r}}.
\end{equation}
Here the coupling constant $g_d = 3^{1.5} g$ to the third harmonic to the qubit as well as its photon number $N_d$ were introduced. Also we used the input power $p_\text{in}$ in dBm, a coupling capacitance $C_c = 5$~fF between the input transmission line and the resonator, a total resonator capacitance of $C_r = 0.4$~pF, and a wave impedance $Z$ of 50~Ohm. The latter values are found by an electromagnetic finite element simulation of the sample design.

In comparing Figs.~\ref{Fig:spec} and \ref{Fig:Calc}, we achieve a very good agreement between our experimental findings and the simulation. Notably, we can accurately relate the applied driving power to the driving amplitude $\Omega_d$ using the observed level shift. By the observed similarities of the results with and without the second order correction terms on coupling and dissipation in the respective regions of positive and negative bias, we also demonstrate that a far detuned driving signal with not too large amplitude will mainly effect the level splitting. We conclude that although a certain amount of absorption is required to introduce an dynamical Zeeman shift to the system, still because the driving signal is far detuned, it may be interpreted as an effective two-level system with the qubit mainly in the ground state. The fact that no adjustment of the dissipative rates, i.e. the spectroscopy line width, is required indicates that the strong microwave tone does not introduce significant broadening, as similar discussed in Ref.~\onlinecite{Gambetta2006}. Therefore, this method can be applied to dynamically modify the level structure of a superconduting qubit in an experiment, either for a fast tunability or for the correction of unavoidable parameter spread due to the microfabrication.

\begin{center}
\begin{figure}[ht]
\includegraphics[width = 8 cm]{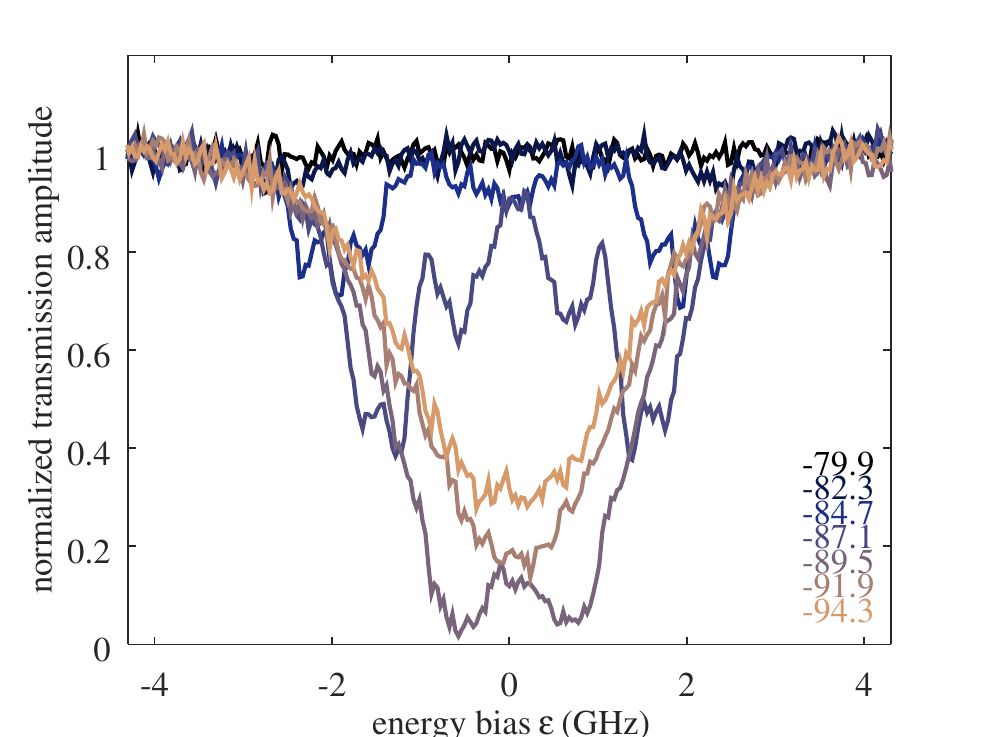}
\caption{(Color online) Transmission amplitude as a function of the energy bias. The different colors represent varying power applied at the third harmonic frequency. Its values are given in dBm at the sample input. With increasing amplitude first the dispersive shift is increased before two instead of one dip appears. These two dips correspond to so-called avoided level crossings: direct interactions between the resonator's fundamental mode and the ac-Zeeman shifted qubit. The amplitude of these dips is decreased as they are shifted away from the degeneracy point ($\varepsilon =0$) by even stronger driving because the qubit coupling scales as $\propto \Delta/\omega_q$. } \label{Fig:toresonance}
\end{figure}
\end{center}

We show a possible application of such a dynamical tuning of the qubit by shifting its energy level splitting into resonance with the fundamental mode of the resonator. Accordingly, we plotted curves of the change in transmission amplitude of the probe signal by the dispersive shift of the resonator that is induced by the qubit while additionally adjusting the driving amplitude $\Omega_d$ in Fig.~\ref{Fig:toresonance}. Because the detuning between resonator and qubit gets smaller with larger driving amplitude, the maximal dispersive shift at zero bias is increased till the resonance condition is met. There the characteristic dip is split into two in the shown transmission amplitude. Plotting the transmission phase (not shown) reveals the two steep phase jumps that result from the double avoided level crossings, compare for example to the results in Ref.~\cite{Oelsner2010} that are also achieved in an intermediate coupling regime. However, for large driving amplitudes several additional resonances  are observed corresponding, for example, to multi-photon resonances and direct interactions with the dressed states. In the experiment, these multiple interactions with the fundamental mode are each observed as dips and phase jumps, making especially the plots of the phase curves rather complicated. Therefore, we conclude, that the dynamical shift of the qubit levels when used for tuning the energy levels should be moderate to limit multiphoton interactions.

\section*{Conclusion}
We demonstrated that a strong off-resonant driving signal can be used to introduce a radiative shift of the energy levels of a superconducting qubit. We showed that the dressed qubit can still be understood as quantum two-level system and standard techniques as coupling to a cavity as well as spectroscopy of the shifted qubit energy levels can be straightforwardly performed. We developed a theoretical model that is based on mapping of corrections found in both, the rotating and counter-rotating frame, back to the laboratory frame. The method is similar as the one used to account for the Bloch-Siegert shift in a rotating frame \cite{FornDiaz2010,Naderi2011} and gives a level shift that coincides with the one found by Floquet theory \cite{Rosenbusch2009}. Furthermore, it enables the description of the driving's influence to the dipole moment of the qubit and the therewith introduced changes in coupling constants and dissipative rates. Although we make use of the strong detuning of the drive in our theory, the calculations can be carried out to any desired precision by including higher order terms. Additionally, our method allows to increase the accuracy of such methods that pick a certain rotating frame for the description of interactions with special dressed states, as for example in Refs.~\onlinecite{Oelsner2013,Shevchenko2014,Koski2018} by including corrections from fast oscillating terms. We also like to note that our work can be extended to account for multi-photon interactions as well as to gain a higher accuracy for stronger driving in the description of certain experimental results.

We believe that for particular applications the method of tuning the energy levels by an ac drive can help to overcome the still not resolved issues of parameter spread and reproducibility in micro-fabricated superconducting quantum circuits. Furthermore, compared to tuning energy levels with low frequency fields, our proposal does not introduce additional low frequency noise and the modification of the dissipative rates is deterministic.

\begin{acknowledgments}
The authors thank Mikhail Fistul for fruitful discussions. We acknowledge the support of Birger Steinbach and Katrin Pippardt at the Competence Center Micro and Nanotechnologies of the Leibniz IPHT in fabricating the sample. This work has received funding from the European Union's Horizon 2020 research and innovation programme under grant agreement No 863313 (SUPERGALAX).
\end{acknowledgments}

\appendix%-----------------------------------------------------------

\section{Approximated Hamiltonian}
\label{App}
To analyze the influence of the strong driving signal, we split the system's Hamiltonian \eqref{Eq:Ham_full} into three parts. The first describes the off-diagonal interaction of the qubit with the driving signal
\begin{equation}\label{Eq:Ham1}
\frac{H^{(1)}}{\hbar} = \frac{\omega_q}{2} \sigma_z + \OmD_d \cos \omega_d t \sigma_x.
\end{equation}
In the second term we summarize the remaining interactions
\begin{equation}\label{Eq:Ham2}
\frac{H^{(2)}}{\hbar} = \left[\fe + \Ome_d\cos \omega_d t\right] \sigma_z + \fD \sigma_x,
\end{equation}
with $f = \Omega_s \cos\omega_s t + g (a^\dagger + a)$. The third term contains solely fundamental mode terms that are not influenced by the transformation $U$. As mentioned above the hat down denotes a multiplication with $\varepsilon/\hbar \omega_d$.

We restrict ourselves to exemplarily demonstrate the transformation of Hamiltonian \eqref{Eq:Ham1}. The second term transforms accordingly. In the rotating frame found by the unitary transformation with $U_t = e^{i\omega_d t \sigma_z/2}$ Hamiltonian $\eqref{Eq:Ham1}$ reads
\begin{equation}\label{Eq:deltam}
\begin{split}
\frac{H^{(1)}_{-}}{\hbar} =& \frac{\delta_-}{2} \sigma_z +\frac{\OmD_d}{2} \left( \sigma_x + \left[\sigma_+ e^{2i\omega_d t} + \hc \right] \right).
\end{split}
\end{equation}
Here  h.c. denotes hermitian conjugate. %Also, we abbreviated $f_x= \hbar\Omega_s^\prime \cos \omega_s t + \hbar g^\prime$ and $f_z = \hbar \Omega_d^* \cos \omega_d t+\hbar\Omega_s^* \cos \omega_s t + \hbar g^*$ to shorten the equations.
The unitary transformation $U_- = e^{i \theta_- \sigma_y} $ with the rotation angle $\tan 2\theta_- = \OmD/\delta_-$ diagonalizes the stationary part of Hamiltonian \eqref{Eq:deltam}. Because we assume a large detuning $\delta_-$ compared to the Rabi frequency $\OmD_d$, we can expand
\begin{equation}
U_-\approx 1 + i \frac{\OmD_d}{2\delta_-} \sigma_y - \frac{\OmD_d^2}{8\delta_-^2} - i\frac{3\Omega_d^3}{16\delta_-^3}\sigma_y
\end{equation}
and keep only terms of the Hamiltonian to second order in $\OmD_d/\delta_-$. The relevant terms of the transformation are
\begin{equation}
\begin{split}
U_- H_-U_-^\dagger \approx& H_- +  \frac{\OmD_d}{2\delta_-}\left( 1 -\frac{\OmD_d^2}{2\delta_-^2}\right) i\left[ \sigma_y, H_- \right] \\
&+ \frac{\OmD_d^2}{4\delta_-} \left( \sigma_y H_- \sigma_y - H_- \right).
\end{split}
\end{equation}
Applying this to the qubit operators we find
\begin{equation}
\begin{split}
U_- \sigma_z U_-^\dagger \approx &  \left(1-\frac{\OmD_d^2}{2\delta_-^2} \right) \sigma_z - \frac{\OmD_d}{\delta_-}\left(1-\frac{\OmD_d^2}{2\delta_-^2} \right) \sigma_x,\\
U_- \sigma_x U_-^\dagger \approx &  \left(1-\frac{\OmD_d^2}{2\delta_-^2} \right) \sigma_x + \frac{\OmD_d}{\delta_-}\left(1-\frac{\OmD_d^2}{2\delta_-^2} \right) \sigma_z ,\\
U_- \sigma_\pm U_-^\dagger \approx & \sigma_\pm + \frac{\OmD_d}{2\delta_-} \left(1-\frac{\OmD_d^2}{2\delta_-^2} \right) \sigma_z  -\frac{\OmD_d^2}{4\delta_-^2} \sigma_x.
\end{split}
\end{equation}
The transformation removes the term $\OmD_d \sigma_x/2 $ from Hamiltonian \eqref{Eq:Ham1} and produces the dynamical shift $\propto \OmD_d^2\sigma_z/\delta_- $ in the new reading Hamiltonian
\begin{equation}\begin{split}
\frac{\widetilde{H}^{(1)}_-}{\hbar} &= \left( \frac{\delta_-}{2} + \frac{\OmD_d^2}{4\delta_-}\right)  \sigma_z + \frac{\OmD_d}{2}\left( \sigma_+ e^{2i\omega_d t} +\hc\right)\\
& + \frac{\OmD_d^2}{2\delta_-} \cos 2\omega_d t \sigma_z - \frac{\OmD_d^3}{4\delta_-^2} \cos2\omega_d t \sigma_x.
\end{split}\end{equation}
Similarly the Bloch-Siegert shift can be calculated by writing above Hamiltonian in a system rotating counter clockwise with the same frequency $\omega_d$ in relation to the laboratory frame. This is achieved by $\widetilde{H}_+ = U_t^\dagger U_t^\dagger \widetilde{H}_- U_t U_t$ and we find
\begin{equation}\begin{split}
\frac{\widetilde{H}^{(1)}_+}{\hbar} &= \left( \frac{\delta_+}{2} + \frac{\OmD_d^2}{4\delta_-}\right)\sigma_z +\frac{\OmD_d}{2} \left(1 - \frac{\OmD_d^2}{4\delta_-^2} \right) \sigma_x\\
&- \frac{\OmD_d^3}{8\delta_-^2} \left( e^{-4i \omega_d t} \sigma_+ + \hc   \right)  + \frac{\OmD_d^2}{2\delta_-} \cos 2\omega_d t\sigma_z.
\end{split}\end{equation}
Similar as before the Hamiltonian is rotated by a small angle $\tan 2\theta_+ = \Omega_d^\prime/\delta_+ $ allowing an expansion of the transformed Hamiltonian to second order in $\Omega_d^\prime/\delta_+$.
\begin{equation}\begin{split}
\frac{\widetilde{H}^{(1)}_+}{\hbar} &= \left( \frac{\delta_+}{2} + \frac{\OmD_d^2}{4\delta_-} + \frac{\OmD_d^2}{4\delta_+} \right)\sigma_z - \left( \frac{\OmD_d^3}{8\delta_-^2} + \frac{\OmD_d^3}{4\delta_- \delta_+}\right)\sigma_x  \\
&- \frac{\OmD_d^3}{8 \delta_-^2} \left( e^{-4i\omega_d t} \sigma_+ + \hc\right) + \frac{\OmD_d^2}{2\delta_-} \cos 2\omega_d t \sigma_z \\
&- \frac{\OmD_d^3}{2\delta_- \delta_+} \cos 2\omega_d t \sigma_x.
\end{split}\end{equation}
Further on we will transform back to the laboratory frame and neglect terms oscillating with multiples of frequency $\omega_d$ and all orders higher than two of $\OmD_d/\delta_\pm$.
\begin{equation}\label{Eq:finalHam1}\begin{split}
\frac{\widehat{H}^{(1)}}{\hbar} &= \left( \frac{\omega_q}{2} + \frac{\OmD_d^2}{4\delta_-} + \frac{\OmD_d^2}{4\delta_+} \right)\sigma_z \\
& -  \frac{\OmD_d^3}{8\delta_-^2} \left( \sigma_+ e^{i\omega_dt} + \hc \right)  - \frac{\OmD_d^3}{4\delta_- \delta_+}  \cos \omega_d t \sigma_x.
\end{split}\end{equation}
With the approximated Hamiltonian above we transformed the strong driving of the qubit into an effective shift of its energy levels. The remaining oscillating terms result from the interplay of the oscillation in the rotating and counter-rotating frame. %Still there amplitude is small for large detunings. Especially in our experiment we can approximate $\omega_q \approx \omega_d/2$ which results in a scaling of the already small assumed driving amplitudes with $\OmD_d^2/\omega_d^2$. Therefore we neglect this remaining oscillations.

The same procedure is applied to Hamiltonian \eqref{Eq:Ham2} resulting in a rescaling of the coupling constants to the fundamental mode and the spectroscopic signal described by the Hamiltonian
\begin{equation}\label{Eq:finalHam2v1}
\begin{split}
\frac{\widehat{H}^{(2)}}{\hbar} =&\Ome_d \left(1-\frac{\OmD_d^2}{2\delta_-^2} -\frac{\OmD_d^2}{2\delta_+^2} +\frac{\OmD_d^2}{2 \delta_-\delta_+} \right) \cos \omega_d t \sigma_z \\
& - \frac{\Ome_d}{\OmD_d}\omega_{ac} \sigma_x + \fe  \left(1-\frac{\OmD_d^2}{2\delta_-^2} -\frac{\OmD_d^2}{2\delta_+^2}\right)\sigma_z \\
&+ \fe  \left[ \frac{\OmD_d}{\delta_+} \sigma_+ e^{i\omega_d t} + \frac{\OmD_d}{\delta_-} \sigma_+ e^{-i\omega_d t} + \hc \right]  \\
&+ \fD \left( \left[ \sigma_x - \frac{\omega_{ac}^2}{\OmD_d^2} \sigma_x  \right]  + \frac{2\omega_{ac}}{\OmD_d} \cos \omega_d t \sigma_z  \right).
\end{split}
\end{equation}
Above, as before, we neglected terms with orders higher than two of $\Omega_d/\delta_\pm$ with any combination of the subscript plus and minus sign, as well as oscillations with multiples of frequencies $\omega_d$ (RWA). Still, we emphasizes that keeping the latter would allow to describe interaction processes with multiple photons of the driving signal and the qubit as well as the fundamental mode and spectroscopic signal. The modifications of Hamiltonian \eqref{Eq:Ham2} as found in \eqref{Eq:finalHam2v1} result from the additional dipole moment of the qubit induced by the driving signal $\Omega_d$. It leads to a reduction of the coupling constants given by the upper line, the second term in the second line, as well as the first term in the brackets of the lower line of \eqref{Eq:finalHam2v1}. These corrections are all second order in $\Omega_d/\delta_\pm$. The diagonal coupling of the qubit to the driving signal additionally produces a stationary transition rate between the modified qubit states as given by the first term in the second line of \eqref{Eq:finalHam2v1}. Other interesting terms are found in the third line of the same equation. They describe interactions between the qubit and the fundamental mode as well as the spectroscopic signal that additionally require driving photons. They actually allow a theoretical description of processes as the spectroscopy of the dressed levels and dressed state amplification or damping. We note, that the related coupling constant scales in the same way as in Ref.~\onlinecite{Oelsner2013} for large detunings.

Because those interactions are not of interest in the present work, we neglect the remaining terms oscillating with $\omega_d$ in \eqref{Eq:finalHam1} and \eqref{Eq:finalHam2} to find the transformed Hamiltonian
\begin{equation}\label{Eq:finalHam2}
\begin{split}
\widehat{H}=&\hbar\frac{\omega_q+ \omega_{ac}}{2} \sigma_z  - \frac{\varepsilon}{\Delta} \hbar\omega_{ac} \sigma_x \\
+& \hbar\left(\Omega_s \cos\omega_s t + g (a^\dagger + a)\right) \left(\frac{\varepsilon A}{\omega_q} \sigma_z + \frac{\Delta B}{\omega_q} \sigma_x \right).
\end{split}
\end{equation}
Above we used $\Ome/\OmD = \varepsilon/\Delta$ and abbreviated $A = 1-\frac{\OmD_d^2}{2\delta_-^2} -\frac{\OmD_d^2}{2\delta_+^2}$ as well as $B = 1  - \frac{\omega_{ac}^2}{\OmD_d^2}$ to keep the equation in a compact form.

\section{Approximated Dissipation}
\label{App2}
The transformations to include the dynamical Zeeman shift to the qubit will also influence to the relaxation dynamics. In principle, we transform the Lindblad operator \eqref{Eq:Linblad} in the same way as the Hamiltonian
\begin{equation}
\begin{split}
\widehat{L} =& \frac{\Gamma_r}{2} \left( 2 U \sigma_-  \rho  \sigma_+ U^\dagger - \left\{ U \sigma_+ \sigma_- U^\dagger, U \rho U^\dagger \right\}  \right)\\
& +\frac{\gamma_\varphi}{2} \left(U \sigma_z \rho \sigma_z U^\dagger - U \rho U^\dagger \right).
\end{split}
\end{equation}
Here we can insert $1 = U^\dagger U$ in each product of operators and note that $\hat{\rho} = U \rho U^\dagger$ is the density matrix in the modified basis. Therefore, we transform each of the containing operators into the new approximated basis. In performing above calculation, the original Lindblad operator given in \eqref{Eq:Linblad} stays the same in the rotating frame found by $U_t$ because concerning the relaxation in each term the factors from  $\sigma_+$ and $\sigma_-$ compensate each other and the dephasing is invariant under rotation around $\sigma_z$. The transformation $U_-$ produces

\begin{equation}\begin{split}\label{Eq:Diss1}
\widetilde{L}_- &=  \frac{\widetilde{\Gamma}_r}{2} \left( 2\sigma_- \widetilde{\rho}_- \sigma_+ - \left\{ \sigma_+ \sigma_- , \widetilde{\rho}_-\right\} \right) + \frac{\widetilde{\gamma}_\varphi}{2} \left( \sigma_z \widetilde{\rho}_- \sigma_z -\widetilde{\rho}_- \right)\\
&+ \frac{\widetilde{\Gamma}_e}{2} \left( 2\sigma_+ \widetilde{\rho}_- \sigma_- - \left\{ \sigma_- \sigma_+ , \widetilde{\rho}_-\right\} \right)\\
& + \frac{\OmD_d \left[ \Gamma_r - \gamma_\varphi \right]}{2\delta_-}\left( \sigma_- \widetilde{\rho}_- \sigma_z + \hc \right)  \\
& - \frac{\OmD_d \gamma_\varphi}{2\delta_-}  \left( \sigma_z \widetilde{\rho}_- \sigma_- + \hc \right) +\frac{\OmD_d \Gamma_r}{4\delta_-}  \left\{ \sigma_+ + \sigma_- ,  \widetilde{\rho}_-\right\} \\
&-\frac{\OmD_d^2 \left[ \Gamma_r - 2 \gamma_\varphi \right]}{4\delta_-^2}   \left(\sigma_- \widetilde{\rho}_- \sigma_- + \hc \right).
\end{split}\end{equation}
We neglected additional dissipation at the Rabi frequency that might be introduced by the approximate diagonalization $U_-$ because we restrict our discussion to strong off-resonant driving \cite{Ithier2005,Hauss2008a}. Above we abbreviated the respective relaxation, excitation, and decoherence rates in the approximated rotating basis
\begin{equation}\begin{split}
\widetilde{\Gamma}_r &= \Gamma_r -  \frac{\OmD_d^2}{2\delta_-^2} \left[ \Gamma_r - \gamma_\varphi \right] , \widetilde{\Gamma}_e = \frac{\OmD_d^2}{2\delta_-^2}  \gamma_\varphi ,\\
\widetilde{\gamma}_\varphi &= \gamma_\varphi + \frac{\OmD_d^2}{2\delta_-^2} \left[\Gamma_r -2 \gamma_\varphi\right].
\end{split}\end{equation}
Note, these rates are consistent with the results found for example in Ref.~\onlinecite{Shevchenko2014} when expanded to second order for small $\OmD_d$. We neglected all terms with orders larger than two of $\OmD_d/\delta_-$. Additionally, we can safely omit the last two terms in Eq.~\eqref{Eq:Diss1} because they will produce only fast oscillating terms back in the laboratory frame after the final transformation.

We repeat the above procedure in the counter-rotating frame and with the transformation $U_+$. Thereby we note that $\widetilde{\Gamma}_e$ is already second order in $\OmD_d/\delta_-$ so that this terms is not further modified. The remaining terms in the approximate diagonal counter rotating frame read
\begin{equation}\begin{split}\label{Eq:Diss2}
\widehat{L}_+ &=\left(\frac{\widetilde{\Gamma}_e}{2} + \frac{\OmD_d^2}{2\delta_+^2}\right) \left( 2\sigma_+ \widehat{\rho}_+ \sigma_- - \left\{ \sigma_- \!\sigma_+ , \widehat{\rho}_+\right\} \right) \\
&+\!\frac{1}{2}\left( \! \widetilde{\Gamma}_r - \frac{\OmD_d^2}{2\delta_+^2} \! \left[ \widetilde{\Gamma}_r - \widetilde{\gamma}_\varphi \right] \! \right)\!\left( 2\sigma_- \widehat{\rho}_+ \sigma_+ \! - \! \left\{ \sigma_+ \!\sigma_-, \widehat{\rho}_+\right\} \right) \\
& + \! \frac{1}{2}\left(\!\widetilde{\gamma}_\varphi + \frac{\OmD_d^2}{2\delta_+^2}\left[\widetilde{\Gamma}_r - 2\widetilde{\gamma}_\varphi \right]\!\right) \! \left( \sigma_z \widehat{\rho}_+ \sigma_z -\widehat{\rho}_+ \right)\\
& + \! \frac{\OmD_d^2}{2\delta_-\delta_+}\left( \sigma_+ \widehat{\rho}_+ \sigma_+ e^{-2i\omega_d t} + \hc \right) \left[ \Gamma_r - 2\gamma_\varphi \right].
\end{split}\end{equation}

We again neglected all terms that are of higher order than two of $\OmD_d/\delta_\pm$. Additionally we only kept terms that are stationary in the laboratory frame in which the final Lindbladian is given by Eq.~\eqref{Eq:Linbladmod}.
%\begin{equation}\begin{split}
%\widehat{L} &=+\frac{\widehat{\Gamma}_r}{2} \left( 2\sigma_- \widehat{\rho} \sigma_+  - \left\{ \sigma_+ \sigma_-, \widehat{\rho}\right\} \right) \\
%&\frac{\widehat{\Gamma}_e}{2} \left( 2\sigma_+ \widehat{\rho} \sigma_- - \left\{ \sigma_- \sigma_+ , \widehat{\rho}\right\} \right) \\
%& + \frac{\widehat{\gamma}_\varphi}{2} \left( \sigma_z \widehat{\rho} \sigma_z -\widehat{\rho} \right).
%\end{split}\end{equation}
Note that compared to Eq.~\eqref{Eq:Diss2}, we neglected the last line because when calculating expectation values (see Eqns. \eqref{Eq:MaxBloch1} - \eqref{Eq:MaxBloch3}) the order of operators under the trace can be changed resulting in a vanishing of these terms. Therefore the basis transformation modifies the relaxation and decoherence rates and adds excitation to the system with the respective rates given in Eq.~\ref{Eq:modrates}.

\bibliography{lit}{}
\end{document}